\documentstyle[prl,aps,epsf]{revtex}

\newcommand{\req}[1]{(\ref{#1})}
\newcommand{\beq}{\begin{equation}}
\newcommand{\eeq}{\end{equation}}
\newcommand{\beqar}{\begin{eqnarray}}
\newcommand{\eeqar}{\end{eqnarray}}

\begin{document}
\draft
\twocolumn[\hsize\textwidth\columnwidth\hsize\csname
@twocolumnfalse\endcsname
\title{Exact solution of stochastic directed sandpile model}
\author{Morten Kloster$^1$, Sergei Maslov$^2$, Chao Tang$^3$}

\address{$^1$ Department of Physics, Princeton University, Princeton, New
Jersey 08544}
\address{$^2$ Department of Physics, Brookhaven National Laboratory,
Upton, New York 11973}
\address{$^3$ NEC Research Institute, 4 Independence Way, Princeton, New
Jersey 08540}

\date{\today}
\maketitle
\begin{abstract}
We introduce and analytically solve a directed sandpile model with stochastic
toppling rules. The model clearly belongs to a different universality class from
its counterpart with deterministic toppling rules, previously solved by Dhar and
Ramaswamy. The critical exponents are $D_{||}=7/4$, $\tau=10/7$ in two dimensions
and $D_{||}=3/2$, $\tau=4/3$ in one dimension. The upper critical dimension of
the model is three, at which the exponents apart from logarithmic corrections
reach their mean-field values $D_{||}=2$, $\tau=3/2$.
\end{abstract}
]

Numerical and analytical studies of sandpile models of Self-Organized Criticality \cite{BTW}
continue to be a subject of considerable research activity.
In particular, a lot of effort has been recently invested in establishing
the set of universality classes in these systems \cite{recent_uc}. The consensus seems to be
that the universality class of a d-dimensional sandpile model is determined
by answers to the following list of questions:

(i) Is it a critical slope or a critical height model? In other words, does
a site topple when its local slope or its height exceeds a certain threshold value.
Critical height models (e.g. Bak-Tang-Wiesenfeld (BTW) model \cite{BTW}) were studied more
extensively in the past and are in general better understood.

(ii) Is sand redistributed isotropically in a toppling event? According to this
property sandpiles could be classified
as isotropic or directed (anisotropic). The common knowledge is that
it is a relevant parameter, i.e. an arbitrary small anisotropy
of toppling rules usually drives the model to the directed universality
class.

(iii) Finally, it is important if the sand is distributed
deterministically or randomly in each individual toppling.
In a model with deterministic toppling rules the configuration
of the bulk of the sandpile remains unchanged if every single site on the lattice
topples exactly once. This additional symmetry is usually important
for the universality class of the model: e.g. the deterministic
one-dimensional isotropic critical height sandpile (1D BTW) has only trivially
distributed avalanches of fractal dimension 2, while its cousins with additional
randomness in toppling rules, such as the Zaitsev model \cite{zaitsev},
Oslo model \cite{oslo}, Linear Interface Model \cite{PMB}, etc., seem
to belong to the same universality class where avalanches have
a non-integer fractal dimension $D \simeq 2.23$
and a probability distribution with a clean power law exponent $\tau \simeq 1.27$.

Despite much numerical and analytical effort on the original BTW sandpile
model (which is a deterministic/isotropic/critical-height model in the above
classification), its critical exponents in two dimensions still
remain controversial \cite{recent_uc}. The situation is somewhat better for directed
models. Soon after the original BTW sandpile model\cite{BTW}, Dhar and Ramaswamy
introduced and exactly solved in all dimensions its directed cousin -- the Dhar-Ramaswamy
(DR) model \cite{DR}.

Both BTW and DR models have deterministic toppling rules. As far as stochastic models
are concerned there is preciously little analytical results.
Apart from an exact solution of a model, equivalent to the
1D stochastic directed sandpile \cite{maslov_zhang},
stochastic sandpiles were studied only numerically.
In this paper we present an analytical study of the {\it stochastic} directed sandpile model in all
dimensions. Stochastic directed sandpiles were brought to the attention of the community in
two recent papers \cite{dir_stochastic}, reporting numerical studies of several variants of
such models in two dimensions.
During the preparation of this manuscript, there
appeared a closely related preprint by Paczuski and Bassler
\cite{paczuski_bassler} in which an analytical study of the
directed stochastic sandpile model was presented and similar
results were obtained. In particular, using
different analytical arguments they have arrived
at the same set of exponents.

The microscopic rules of the stochastic directed sandpile model
that we selected to study are
closely related to those of the DR 
model \cite{DR}. These rules are modified in the
spirit of a stochastic isotropic sandpile model known as the Manna model \cite{manna}.
It is easier to define our rules in
two dimensions, while generalization to higher dimensions is straightforward.
A stable configuration of our model is specified by the integer height of
the sandpile $z(x_1,x_2)\leq 1$ at each point of a 2D square lattice.
The lattice has open boundary conditions along the diagonal coordinate,
$x_{||}=x_1+x_2$, and periodic boundary conditions in the transversal direction
$x_{\perp}=x_1-x_2$. The sand is added randomly at the line with $x_{||}=0$
and falls off the edge at $x_{||}=L_{||}$. The difference between our model and
the DR model lies in toppling rules. In both cases, once the
height at any given site exceeds one, this site becomes unstable and
loses two grains of sand to its nearest neighbors
in the direction of increasing $x_{||}$.
However, while in the DR model each of these two neighbors
gets exactly one grain of sand, in our stochastic variant the decision where to
move any particular grain is done independently for each grain. In other
words, with probability $1/4$ both grains end up on the left neighbor, with
probability $1/4$ they go to the right neighbor, and only with probability $1/2$
will each neighbor get one grain as in the DR model. Obviously, on average
each neighbor gets one grain, yet the additional stochastic element
in these rules drives the model away from the universality class
of the DR model \cite{dir_stochastic}.
It is easy to see that unlike the deterministic rules of the DR model,
the new stochastic rules allow for multiple topplings of some sites within
one avalanche. Indeed,
let us consider an example where the first active site of the avalanche toppled
one grain of sand to each of its two neighbors. Let us further
assume that these neighbors both had $z=1$ so that they both toppled
and by chance distributed all four of the resulting grains of sand to the
same site in the next layer. This site has received four grains of sand and
is guaranteed to topple twice. The numerical simulations
\cite{dir_stochastic} confirm the existence of multiple topplings in other variants 
of a stochastic directed sandpile model.

In order to get an analytical handle on the properties of our model we
employ the same trick that was used by one of us to solve the
1D directed stochastic sandpile model \cite{maslov_zhang}.
Due to the Abelian nature of the model,
we can change the order in which topplings are performed without changing the
outcome. It is convenient to do topplings layer by layer. This means that we
topple any given site as many times as necessary to make it stable before
moving on to the next unstable site, and we topple all unstable sites in one layer
(a given $x_{||}$) before toppling any site in the next layer.
Let us concentrate on a site
with coordinates $x_{||}$ and $x_{\perp}$ immediately after we have finished
with topplings in the ($x_{||}-1$)-th layer. Assume that two of its neighbors with coordinates 
$x_{||}-1$ and $x_{\perp} \pm 1$ have toppled, respectively, $n_1=n(x_{||}-1, x_{\perp}-1)$
and $n_2=n(x_{||}-1, x_{\perp}+1)$ times. The average
number of grains of sand that our selected site would receive from the
previous layer is $(n_1+n_2)$. In the DR model there are no fluctuations
around this average. Also, due to the absence of multiple topplings
in this deterministic directed model,
$n_1$ and $n_2$ can be only $0$ or $1$. Therefore, in the DR model
a site can receive either $n_1+n_2=2$ grains, in which case it is guaranteed
to topple exactly once, or $n_1+n_2=1$, in which case it can topple with probability
$1/2$ (i.e. it topples if it had $z=1$ before the transfer and remains
stable if it had $z=0$).
From this one can show \cite{DR} that in the DR model the set of sites which topple at each layer
form an interval with no holes inside. The size of this interval as a function
of the layer number $x_{||}$ performs an ordinary random walk. 

In the stochastic model the relation between the number of topplings in
two subsequent layers is more complicated. Let us concentrate on the
behavior of the total number of topplings $N(x_{||})=
\sum_{x_{\perp}} n(x_{||}, x_{\perp})$ in a given layer $x_{||}$.
This number fixes the number of grains of sand transferred from
the layer $x_{||}$ to the next layer as $2 N(x_{||})$.
It is easy to see that a site
which has received an even number $2k$ of grains of sand from a previous layer
will always topple {\it exactly} $k$ times and, therefore, will
transfer the same $2k$
grains of sand to the layer directly below it.
That means that as far as $N(x_{||})$ is concerned, such sites behave
in a completely passive manner, i.e. they do not lead to a decrease or an increase
of the total number of topplings $N(x_{||})$ from layer to layer.
On the other hand, any site which
received an odd number $2k+1$ of grains of sand from a previous layer
has equal chance to topple $k$ times (if it had $z=0$ before the  transfer) 
or $k+1$ times (if it had $z=1$). In the former case this
site would decrease the grain flow $2N(x_{||})$ by one, while in the latter -- increase by 1.
Let us call any site which has received an odd number of grains of sand from the
previous layer an {\it active site}. The equation, which is the central result
of this work, relating the change in the total number of topplings from
layer to layer to the number of active sites $N_a (x_{||})$ in a given
layer is $N(x_{||})=N(x_{||}-1)+\frac{1}{2}\sum_{a=1}^{N_a(x_{||})} \xi_a$, where
all $\xi_a$ are $-1$ or $+1$ with equal probability and independent of each other.
These random numbers correspond to whether each of the $N_a(x_{||})$ active
sites had the height $z=0$ or $z=1$ before the avalanche started. 
It is straightforward to demonstrate that, as in the DR model, in the steady state of
the directed stochastic model all possible stable configurations of $z$ are equally represented,
and, therefore, there are no correlations between the heights at different sites, and
each height is equally likely to have $z=0$ or $z=1$. 
It is more convenient to rewrite the above equation
in a continuous notation, which works as long as $N_a (x_{||})>>1$ :
\beq
{d N (x_{||}) \over d x_{||}}=\frac{1}{2}\sqrt{N_a (x_{||})} \ \eta(x_{||}) .
\label{main}
\eeq
Here $\eta(t)$ is a standard Gaussian variable with zero mean and
the standard deviation equal to unity. This equation describes an
unbiased random walk $N(x_{||})$ vs $x_{||}$ with a variable step, given
by $\frac{1}{2}\sqrt{N_a(x_{||})}$. A random walk (an avalanche) starts with
$N(0)=1$ and ends at $x_{||}$ when $N(x_{||}) \leq 0$ for the
first time. Let us assume that both $N(x_{||})$ and
$N_a(x_{||})$ in a surviving avalanche scale with $x_{||}$
with the exponents $\alpha$ and $\alpha_a$, respectively.
Alternatively one can say that they are related to each other by
$N_a \sim N^{\alpha_a/\alpha}$. Plugging this relation into
Eq. \req{main}, after easy algebra we get the exponent relation
\beq
\alpha={1 + \alpha_a \over 2} .
\label{alpha}
\eeq
In addition to this exponent relation, the mapping of 
the avalanche process onto a random
walk immediately gives us the power law $\tau_{||}$ in the
probability distribution of avalanche sizes.
Indeed, an avalanche ends when the random walk described by 
Eq. \req{main} first enters the $N(x_{||}) \leq 0$
semi-axis, and it is a standard result of the theory of 
stochastic processes that the distribution of first returns 
of a generalized random walk has an exponent $\tau_{||}=1+\alpha$. 
This agrees with the well-known exponent relation $\tau_{||}=D_{||}$
valid for a general directed sandpile model.
Indeed, the fractal dimension $D_{||}$, defined by
$\sum _{i=1}^{x_{||}} N(i)=S \sim x_{||}^{D_{||}}$,
is obviously related to $\alpha$ through $D_{||}=1+\alpha$.

The Eq.~\req{main} applies equally well to the DR and the stochastic directed
sandpile models. The difference between these two models lies
only in the scaling of the number of active sites with $x_{||}$.
As was explained above, in the 2D DR model the only two active sites
lie at the edge of an interval of toppled sites. Indeed, only these
sites get 1 grain of sand, while the rest get either 0 or 2.
Therefore, in the 2D DR model $N_a(x_{||})=2$ is just a constant, $\alpha_a=0$,
and the Eq. \req{main} describes an ordinary random walk, in which
$N(x_{||}) \sim x_{||}^{\alpha}= x_{||}^{1/2}$.
The introduction of a stochastic element in particle redistribution
dramatically changes the number of active sites at any given layer
of the avalanche. Indeed, when grains are distributed independently,
any site which has at least one {\it toppled neighbor} in the previous layer 
is equally likely to receive an even or odd number of grains of sand, and 
therefore, it has a probability $1/2$ of becoming active. 
Thus, in the stochastic model the exponent $\alpha_a$ defines how the number
of {\it distinct sites} that topple at least once, scales with the
layer number $x_{||}$. The difference between exponents
$\alpha$ and $\alpha_a$ comes solely from the existence of multiple
topplings. These two exponents have to obey the inequality
$\alpha \geq \alpha_a$, and their difference $\alpha - \alpha_a$ determines
how the average number of topplings $n_{\rm top}(x_{||})$ at a given site 
in $x_{||}$-th layer scales with $x_{||}$:
$n_{\rm top} \sim N/(2N_a) \sim x_{||}^{\alpha-\alpha_a}$.

We proceed with an argument that in the 2D directed stochastic model
$\alpha_a=1/2$, and, therefore, by the virtue of Eq. \req{alpha}
$\alpha=3/4$. It is a straightforward task to determine the
{\it average} number of topplings $\langle n(x_{||}, x_{\perp}) \rangle$
at a given site $x_{||}, x_{\perp}$,
where the average is performed over the whole ensemble of avalanches, so that 
avalanches that die out before reaching this site, contribute 0 to the 
average. As was noted in \cite{DR}, due to the conservation of sand and the 
stationarity of the sandpile $\langle n(x_{||}, x_{\perp}) \rangle$
has to satisfy the diffusion equation with a source:
\beq
{\partial \langle n(x_{||}, x_{\perp}) \rangle \over \partial x_{||}}=
{1 \over 2} {\partial^2 \langle n(x_{||}, x_{\perp}) \rangle \over \partial x_{\perp}^2}
+\delta(x_{||})\delta(x_{\perp}).
\eeq
This average balance equation is also exact for our stochastic model, where  
it proves that, like in the DR model, sites that topple at 
least once are spread over the interval
of length $\Delta x_{\perp} \sim x_{||}^{1/2}$ in the $x_{||}$ layer.
In the DR model these sites form a dense interval with no holes, and, 
therefore, their number is known to
scale exactly as $x_{||}^{1/2}$. The situation is somewhat less obvious
in the stochastic directed model, where the set of 
toppled sites can have holes. However, one can argue that
these holes would mostly be concentrated near the boundaries of
the avalanche in any given layer, while the majority
of toppled sites in the center would form a compact interval.
Indeed, as will be confirmed later, the 2D stochastic directed model
is characterized by multiple topplings, where a site
at a layer $x_{||}$ would typically topple 
$n_{\rm top}(x_{||}) \sim x_{||}^{1/4}$
times within one avalanche. Since any of the $2n_{\rm top}$ grains 
is equally likely to go to each of the two nearest neighbors 
in the next layer, the situation where one of these neighbors 
would receive less than two grains is exponentially unlikely
for large $n_{\rm top}$. 
But once a site has received two or more grains of sand it
is guaranteed to topple at least once.
Therefore, both nearest neighbors down the slope from the
site which toppled many times would most likely
topple at least once. In other words, the creation of a new hole (a region
free of topplings) is exponentially suppressed near sites which 
themselves toppled many times. This means that for a sufficiently large $x_{||}$
the majority of sites (especially those close to the center of the avalanche region)
will belong to a hole-free region. Since the size of the interval 
covered by an avalanche scales as $x_{||}^{1/2}$,
the number of active sites should also scale as $N_a \sim x_{||}^{1/2}$.
From $\alpha_a=1/2$ with the help of Eq. \req{main} one gets 
$\alpha=3/4$, $\tau_{||}=D_{||}=7/4$, and $\tau=1+(\tau_{||}-1)/D_{||}=10/7$.
These results are in a nice agreement both with previous numerical simulations
of various versions of stochastic directed sandpile model in two dimensions
\cite{dir_stochastic} and with our own simulations of the model.
In Fig. 1 we present the results of our simulations for the effective
exponents $\alpha=d \log N/d \log x_{||}$ and $\alpha_a=d \log N_a/d
\log x_{||}$ as a function of $x_{||}$.
The numerical exponent $\alpha$ nicely agrees
with the analytical results.  
The exponent $\alpha_a$ is less clean
due to the presence of holes near the boundary of the avalanche
region. The exponent seems first to overshoot to a value of almost $0.6$ but then 
goes down so that in the end of the range of our simulations, $x_{||} \sim 30000$, 
it is consistent with our theoretical prediction $\alpha_a=1/2$.

Unlike its deterministic cousin, the stochastic directed sandpile model
exhibits a non-trivial scaling even in one dimension. The 1D deterministic
directed sandpile model is trivial in the sense that there is just one 
SOC configuration and the addition of a grain of sand always results
in an avalanche of $L_{||}$ topplings in which this grain is transported
and discarded at the open boundary of the system. A 1D stochastic 
directed sandpile model can have several variants of simple microscopic toppling rules.
In one variant, which is essentially identical to the model studied by one
of us in \cite{maslov_zhang}, once a height at a given site exceeds one,
either one or two grains are transferred to the nearest neighbor down the slope.
It is easy to see (for details see \cite{maslov_zhang}) that this model
is equivalent to a 1D random walk so that $N_a={\rm const}$, while 
the typical number of topplings $N$ scales as a function of $x_{||}=x$ 
as $N(x) \sim x^{1/2}$. The distribution of avalanche spatial length 
in this model has an exponent $\tau_{||}=3/2$,
while that of avalanche volume -- $\tau=4/3$. 
In another variant of the 1D stochastic toppling rules an unstable
site always loses two grains of sand and each of these grains with equal probability
goes to the nearest or next-nearest neighbors down the slope.  

As in the DR model the upper critical dimension for the stochastic
directed sandpile model is $d_u=3$. In this dimension the expected number
of topplings at each site in a layer $x_{||}$ grows only logarithmically with $x_{||}$.
Therefore, $\alpha=\alpha_a$ apart from the logarithmic corrections. From
Eq. \req{main} in this case we get $\alpha=(1+\alpha)/2$, which has the solution 
$\alpha=1$, $\tau_{||}=2$, $\tau=3/2$. This is a standard set of mean-field exponents 
for any branching (avalanche) process in high enough dimension. In Fig.
2 we plot the numerical effective exponents in the 3D stochastic directed
model. They agree well with the mean field values.

In conclusion, we have found an analytic solution of the stochastic directed
sandpile model in any dimension. The main difference of this model from its
deterministic counterpart -- the Dhar-Ramaswamy model -- lies in the fractal dimension
of the set of {\it active} sites, i.e sites that can
add or remove one grain from the overall flow of sand between two subsequent layers.
Whereas in the 2D DR model in any layer there are only two sites at the edges 
of the interval of toppled sites which are active, in the 2D stochastic directed sandpile model
each of approximately $x_{||}^{1/2}$ toppled sites in this interval
has a 1/2 chance of being active in the above sense. This leads to an
increase in the fractal dimension of an avalanche from $D_{||}=3/2$ to 
$D_{||}=7/4$ due to the appearance of multiple topplings. 
The difference between critical properties
of stochastic and deterministic directed models disappears in high dimensions $d \geq 3$, where
multiple topplings in a stochastic directed sandpile become prohibitively
unlikely and all exponents acquire their mean-field values.

\begin{figure}
\narrowtext
\centerline{\epsfxsize=3.375in
\epsffile{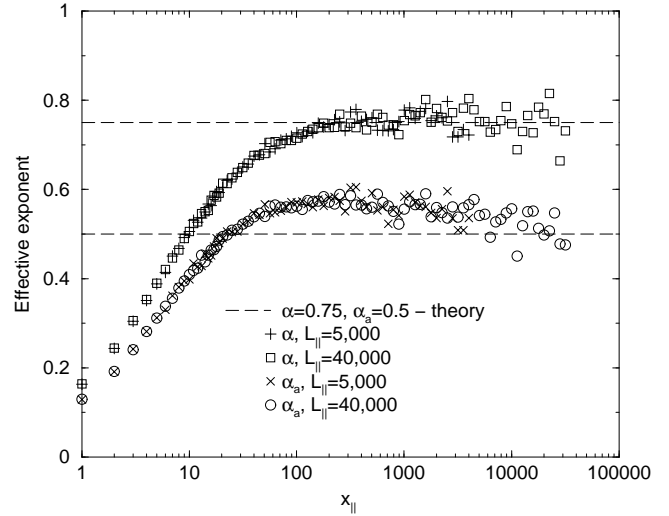}}
\caption{
The scale dependent effective exponents $\alpha=d \log N/d \log
x_{||}$ and $\alpha_a=d \log N_a/d \log x_{||}$ as a function of
$x_{||}$ for the 2D model, and for two longitudinal system sizes
$L_{||}=5000$ and $L_{||}=40000$.}
\end{figure}

\begin{figure}
\narrowtext
\centerline{\epsfxsize=3.0in
\epsffile{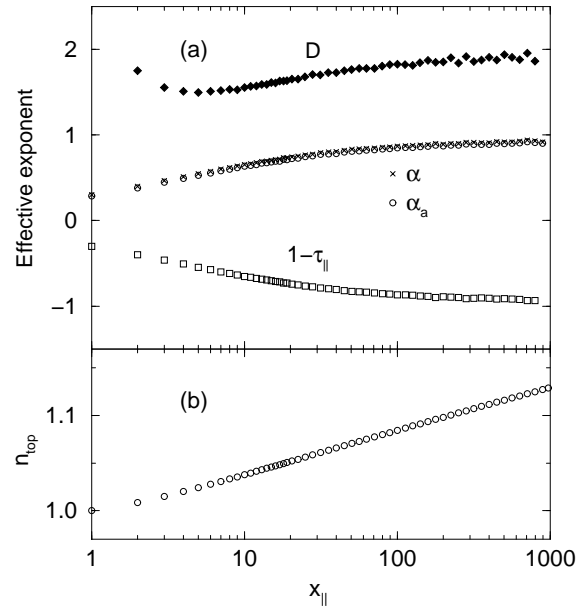}}
\caption{(a) The effective exponents $D, \alpha, \alpha_a$, and $\tau$ as a
function of $x_{||}$ for the 3D model. (b) The expected number of
topplings at each site $n_{top}$ as a function of $x_{||}$. Note the
logarithmical dependence on $x_{||}$.}
\end{figure}

\end{document}